\documentclass[prb,aps,twocolumn,amsmath,amssymb,superscriptaddress]{revtex4-1}
\usepackage{epsfig, graphicx,graphics,amsmath,amssymb,float}
\usepackage[T1]{fontenc}
\usepackage[latin9]{inputenc}
\usepackage{amsmath}
\usepackage{amssymb}
\usepackage{xcolor}
\usepackage{amscd}
\usepackage{bm}
\usepackage{psfrag}

\usepackage{bbm} 

\usepackage[caption=false]{subfig}
\usepackage{subfig}
\usepackage{color}
\usepackage[bookmarks=true,colorlinks,linkcolor=blue,urlcolor=blue,citecolor=blue]{hyperref}
\usepackage{soul}

\begin{document}

\title{Odd-frequency superconductivity in dilute magnetic superconductors}

\author{Fl\'avio L. N. Santos}
\affiliation{Departamento de F\'isica, Universidade Federal de Minas Gerais, Caixa Postal 702,  Belo Horizonte, Minas Gerais, 30123-970, Brazil}
\affiliation{Universit\'e Paris-Saclay, CNRS, Laboratoire de Physique des Solides, 91405, Orsay, France}
\author{Vivien Perrin}
\affiliation{Universit\'e Paris-Saclay, CNRS, Laboratoire de Physique des Solides, 91405, Orsay, France}
\author{Fran\c cois Jamet}
\affiliation{Universit\'e Paris-Saclay, CNRS, Laboratoire de Physique des Solides, 91405, Orsay, France}
\author{Marcello Civelli}
\affiliation{Universit\'e Paris-Saclay, CNRS, Laboratoire de Physique des Solides, 91405, Orsay, France}
\author{Pascal Simon}
\affiliation{Universit\'e Paris-Saclay, CNRS, Laboratoire de Physique des Solides, 91405, Orsay, France}
\author{Maria C. O. Aguiar}
\affiliation{Departamento de F\'isica, Universidade Federal de Minas Gerais, Caixa Postal 702, Belo Horizonte, Minas Gerais, 30123-970,  Brazil}
\affiliation{Universit\'e Paris-Saclay, CNRS, Laboratoire de Physique des Solides, 91405, Orsay, France}
\author{Eduardo Miranda}
\affiliation{Gleb Wataghin Institute of Physics, The University of Campinas (Unicamp), 13083-859 Campinas, SP, Brazil}
\author{Marcelo J. Rozenberg}
\affiliation{Universit\'e Paris-Saclay, CNRS, Laboratoire de Physique des Solides, 91405, Orsay, France}

\begin{abstract}

  We show that dilute magnetic impurities in a conventional superconductor give origin to an odd-frequency
  component of superconductivity, manifesting itself in Yu-Shiba-Rusinov bands forming within the bulk
  superconducting gap. Our results are obtained in a general model solved within the dynamical mean field
  theory. By exploiting a disorder analysis and the limit to a single impurity, we are able to provide general
  expressions that can be used to extract explicitly the odd-frequency superconducting function from scanning
  tunneling measurements.   

\end{abstract}

\maketitle

\section{Introduction}
In superconductors, Fermi statistics imposes that the superconducting pairing function is antisymmetric
under exchange of the two electrons forming the Cooper pairs. The pairing function must therefore change
sign under the exchange of the quantum numbers labeling the two electrons such as position, time,
orbital index, spin, etc. In the most conventional single band spin-singlet superconductivity\cite{Bardeen1957}, it is the spin component of the pairing function which is antisymmetric, while the space part is symmetric (e.g.~$s$-wave). In spin-triplet superconductivity, often advocated in the so-called ferromagnetic superconductors\cite{Ran2019}, such as UGe$_2$, URhGe, and UCoGe, the spin component is symmetric while the space component is antisymmetric (e.g.~$p$-wave). A sign change may occur also in other degrees of freedom.
More than 40 years ago, Berezinskii proposed that the antisymmetric contribution to the pairing function may
derive from the exchange of the electron time coordinates\cite{berezinskii1974new}. He proposed that such a
situation occurs in $^3$He, where the space ($s$-wave) and spin (triplet) components would be symmetric but the time component antisymmetric. In this case, in the space reciprocal to time, the pairing is an odd function of frequency (odd-$\omega$) \cite{Linder,Tanaka2012}.
Since then, odd-$\omega$ pairing was predicted to be quite a general phenomenon in superconducting systems\cite{Balatsky1992,Abrahams1995}, including for example disordered superconductors \cite{Kirkpatrick1991} and heavy-fermion superconductors~\cite{coleman1994odd}.

In more recent years it was realized that an odd-$\omega$ pairing component can arise when superconductivity is induced in ferromagnetic systems by proximity to a conventional superconductor\cite{Bergeret2005}.
In this case the breaking of time-reversal symmetry induced by an effective magnetic field can change the
spin-component of the pairing function from being anti-symmetric to symmetric, favoring the appearance of a
time-antisymmetric component. Such systems offer the advantage of being artificially built and controlled, opening
a path towards applications in the field of spintronic devices\cite{Linder2015,Eschrig2015}. More generally, it
has been shown that an odd-$\omega$ component can indeed arise whenever symmetry breaking occurs, e.g.~the spatial symmetry breaking in non-magnetic junctions \cite{Tanaka2007a,Eschrig2007}. These phenomena have
gained more and more interest with the advent of topological materials, where the competition between
superconductivity and magnetic orders is often a key ingredient.
For instance, proximity effect on a dense chain or wire  of magnetic atoms deposited on  top of a superconductor  gives rise to
unconventional superconductivity, marked by the appearance of Majorana edge states localized at the extremities of the chain~\cite{PergeSci,Ruby2015,Pawlak2016,jeon2017distinguishing,kim2018toward}.  These emergent degrees of
freedom promise to be the
fundamental building blocks in the development of quantum computers~\cite{Nayak2008,Aasen2016}. 
Another interesting system is realized by magnetic islands on the surface of conventional
superconductors 
~\cite{Nakosai2013,Rontynen2015,Li2016b}. Here some experimental signatures of topological superconductivity and chiral
Majorana edge channels have been reported~\cite{Menard2017,Palacio-Morales2019}. 
Understanding the role played by odd-$\omega$ superconductivity in
such Majorana systems and finding what are its experimental signatures, are fundamental open questions~\cite{Cayao2020}.

There has been then a remarkable effort from a large part of the condensed matter community to experimentally
reveal odd-$\omega$ pairing. Proposals include, e.g., measurements of thermopower in
superconductor-quantum-dot-ferromagnet hybrid systems~\cite{hwang2018odd} and the Josephson effect in
superconductor-ferromagnet junctions~\cite{Linder}.
Despite this, the detection of odd-$\omega$ pairing has 
remained a theoretical chimera and only
very recently has its experimental realization apparently been confirmed. 
One recent study has reported odd-$\omega$ superconductivity
at the interface of a topological insulator with a conventional superconductor~\cite{krieger2020proximityinduced}. 
In another recent development, which builds upon earlier theoretical work~\cite{weiss2017odd,kuzmanovski2020odd},
the presence of an odd-$\omega$ component in scanning tunneling spectroscopy (STS) has finally been reported
in a system of a magnetic impurity in contact with a conventional superconductor~\cite{perrin2019unveiling}.

Following the latter study, the goal of the present work is to go beyond the single impurity system, and show that odd-$\omega$ pairing can be induced in a conventional superconductor by the collective
effect of a finite concentration of magnetic impurities. For this purpose, we
consider a general model of magnetic impurity sites embedded within a conventional superconducting lattice
(as portrayed in Fig.~\ref{lattice}): a dilute magnetic superconductor (DMS).

The article is organized as follows. In Sec.~\ref{modelmethod}  we  introduce our model and 
the dynamical mean field theory (DMFT\cite{Rozenberg}) method, which allows us to solve it in a well
controlled infinite dimensional limit. Moreover, DMFT is a mean field theory based on 
Green's functions, thus has the advantage of providing the local spectral functions, which may be directly
observed in STS experiments.
Our results are presented in Sec.~\ref{result1}, 
where show  
the appearance of impurity bands inside the superconducting gap,
the Yu-Shiba-Rusinov bands, which possess an odd-$\omega$ component. Differently from the single impurity
case\cite{kuzmanovski2020odd,perrin2019unveiling},  odd-$\omega$ superconducting
pairing is present for the whole system, for both magnetic and non-magnetic sites, 
making it eventually exploitable in transport and device making.
In Sec.~\ref{3a} 
we develop an impurity-concentration scaling analysis that allows us to derive an explicit expression relating
the odd-$\omega$ superconducting function to the STS local density of states. This enables us to explicitly extract
the odd-$\omega$ superconducting pairing function, which we compare with the exact DMFT solution.
In Sec.~\ref{deriv} we provide a detailed analytic proof of the relations previously derived by
considering the diluted disorder limit of the DMFT solution.
Finally, Sec.~\ref{conclu} provides a summary of our results in order to motivate future experimental investigations
in dilute magnetic superconductors.





\section{Model and Method\label{modelmethod}}

\begin{figure}[t] 
\includegraphics[scale=1.20]{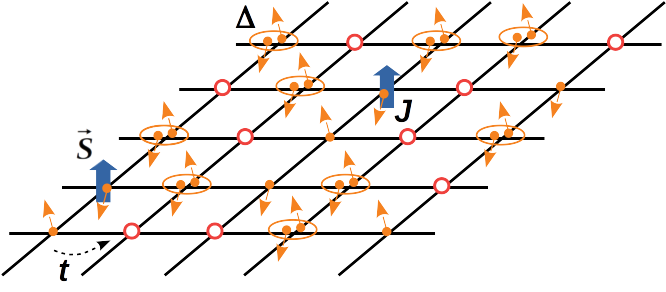}
\caption{Dilute magnetic superconductors: magnetic impurities sites are embedded in a superconducting lattice.
The impurity-site magnetic moment $\vec{S}$ (represented by the blue arrow) interacts with electrons (in orange) via a magnetic coupling $J$, as described by the Hamiltonian in Eq.~(\ref{impclass}).}\label{lattice}
\end{figure}

The DMS model portrayed in Fig.~\ref{lattice} is described by the Hamiltonian
\begin{eqnarray}
 H&=&\sum_{\langle ij\rangle,\alpha}t_{ij}c_{i\alpha}^\dagger c_{j\alpha} +  \sum_i(\Delta_i c_{i\downarrow}c_{i\uparrow}+\Delta_i ^* c_{i\uparrow}^\dagger c_{i\downarrow}^\dagger)\nonumber\\
& & -\sum_{i,\alpha}\mu_i(c_{i\alpha}^\dagger c_{i\alpha})
  + J\sum_{l,\theta,\alpha,\beta}S_l^\theta c_{l\alpha}^\dagger \sigma_{\alpha\beta}^\theta c_{l\beta}.
\label{impclass}
\end{eqnarray}
Here the operator $c_{i,\alpha}^\dagger$ creates an electron with spin projection $\alpha$ at site $i$ and $t_{ij}$ is the  hopping amplitude between 
neighboring sites $\langle ij \rangle$. At random sites $l$ we place classical magnetic moments $S_l^\theta$ (with components $\theta=x,y,z$), which couple to the electrons via an exchange parameter $J$ (here $\sigma^\theta_{\alpha\beta}$
are the Pauli matrices with the spin indices $\alpha, \beta$). 
The $\Delta_i$ term describes superconducting pairing, and we consider $\Delta_i=\Delta$ on non-magnetic sites while $\Delta_i=0$ on magnetic sites. The particle density is fixed by the chemical potential $\mu$, and we fix $\mu_i=\mu$ on non-magnetic sites and $\mu_i=\mu+\delta_\mu$ on magnetic sites, where $\delta_\mu$ is an energy offset that describes potential disorder.
Magnetic sites are uniformly distributed, randomly occurring at a given site with probability $x$  ($0\leq x\leq 1$). 
It was shown in the Supplemental Material of Ref.~\onlinecite{perrin2019unveiling} that spin-orbit coupling (SOC) does not affect the local Green's functions in the case of a single magnetic impurity in the superconductor. We then expect the effect of SOC on the local Green's functions to be negligible in the case of dilute magnetic impurities, $x\ll 1$, and do not consider it in the Hamiltonian.

As mentioned in the introduction, this many-body Hamiltonian can be solved in a well-controlled fashion in the infinite dimensional limit, where the $J$ coupling 
can be treated beyond perturbation theory by means of DMFT\cite{Rozenberg}. 
Within this method the interaction reduces to a purely local term, and the full lattice problem is mapped 
on a quantum impurity model coupled to an effective bath of non-interacting fermions. 
In this case,
we need to consider only the on-site one-particle propagator, which in the superconducting state can be conveniently 
expressed as a 2$\times$2 Nambu matrix related to the Nambu spinor $\psi= (c_\uparrow~c_{\downarrow} ^\dagger)^T$, 
\begin{equation}
   \hat{G}(\omega)= \left(
   \begin{array}{cc}
G_\uparrow(\omega) & F(\omega)\\
F(\omega) & -G_\downarrow^*(-\omega)
   \end{array}\right).\label{matrix22}
\end{equation}
Here $G_\alpha(\omega)$ and $F(\omega)$ are, respectively, the normal 
and anomalous components of the Fourier transform 
on the Matsubara axis of the Nambu Green's function
$\hat{G}(\tau)=-\langle T \psi(\tau)\psi^\dagger(0)\rangle$, 
where $T$ is the time ordering operator for imaginary time $\tau$.

The goal in the DMFT approach is to determine the right hybridization function, or bath, of the quantum impurity, which satisfies 
the DMFT self-consistency condition, that is specified by the original lattice model of Eq.~(\ref{impclass}).
In our problem, however, there are two inequivalent lattice sites, the magnetic and the non-magnetic ones. 
Thus, we must consider two separate quantum impurity problems, one magnetic and the other non-magnetic,
which become coupled through the self-consistency condition.

We introduce some simplifying assumptions, which should not qualitatively change the nature
of the model behavior. First, the classical magnetic moments $S_l$ are assumed to be frozen, 
acting as magnetic spin disorder. 
With this assumption, the DMFT method turns out to be equivalent to the treatment of magnetic disorder within the so-called coherent potential approximation (CPA)\cite{Yonezawa,RevModPhys.46.465,Ziman}.
Second, we assume that the magnetic impurities are fully polarized and keep only the $S^z$ 
component in the Hamiltonian of Eq.~(\ref{impclass}).  
Finally, without loss of physical generality, in the infinite dimensional limit it is convenient 
to adopt a Bethe lattice with hopping $t_{ij}=t/\sqrt{z}$, where $z\to \infty$ is the number of first neighbors of each site, 
whose density of states is a simple semicircle 
$D(\epsilon)=\sqrt{4t^2-\epsilon^2}/(2\pi t^2)$ \cite{Rozenberg}. 
This greatly simplifies the DMFT equations, which can be written as 
the two coupled Green's functions equations at non-magnetic ($nm$) and magnetic ($m$) sites

 \begin{eqnarray}
\hat{G}_{nm}^{-1}(i\omega)&=&i\omega\openone + \mu\tau^z-\Delta\tau^x-t^2\tau^z\hat{G}_{av}\tau^z,   
   \nonumber \\      
   \hat{G}_m^{-1}(i\omega)&=&i\omega\openone + (\mu+\delta_\mu)\tau^z-t^2\tau^z\hat{G}_{av}\tau^z -J\openone.\label{dmfteq22}
 \end{eqnarray}
Here $ \hat{G}_{av}=x\hat{G}_m+(1-x)\hat{G}_{nm}$ is the Green's function averaged over $m$ and $nm$ sites and $\tau^{x,z}$ are Pauli matrices in the Nambu spinor indices.
These equations are solved numerically. 
For definiteness, in the following figures we fix the Hamiltonian parameters
$t=1,~ \mu=-0.05,~ \delta_\mu=-0.5$, $\Delta=-0.1$ and $J=-0.65$. 
The results that we describe next are rather generic, namely, they do not depend on any particular 
fine tuning of parameters. The model parameters are physically reasonable, as the conventional superconductor 
has a relatively small gap and the non-interacting density of states at the chemical potential is featureless.

\section{Results\label{result1}}

\begin{figure}[t] 
 \includegraphics[scale=0.62]{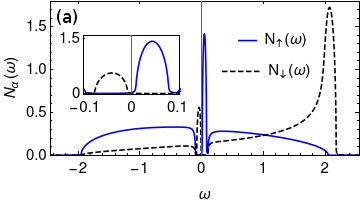}
 \includegraphics[scale=0.65]{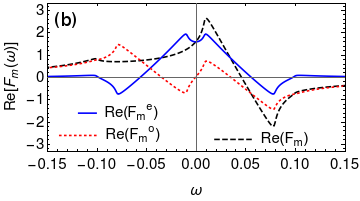}
 \includegraphics[scale=0.65]{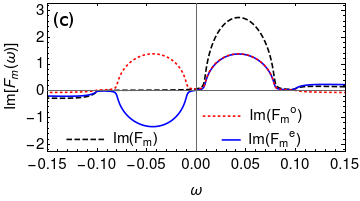}
\caption{Numerical results obtained using the DMFT equations (\ref{dmfteq22}) with $t=1,~ \mu=-0.05,~\delta_\mu=-0.5,~ \Delta=-0.1,~J=-0.65,~ x=0.01$ and broadening $\eta=10^{-3}$. In panel (a) we show the full spectrum for $N_\uparrow$ and $N_\downarrow$;  in the inset we see these functions at low energies, $|\omega|<|\Delta|$. In (b) and (c) we present the real and imaginary components of the superconducting function $F_m(\omega)$, respectively, as well as its even-$\omega$ and odd-$\omega$ components, for low energy.}\label{STM1}
\end{figure}

A single magnetic impurity interacting with the electrons in a superconductor gives origin to electron bound states at the impurity
site known as  Yu-Shiba-Rusinov (YSR) states. This YSR state appears as a sharp resonance within the superconducting gap 
in the spectral density of states, which can be experimentally revealed, e.g., by STS. 
Such a phenomenon is reminiscent of the bound states in 
dilute magnetic semiconductors\cite{Millis}, where the role of the superconducting gap is played by the semi-conductor gap. 
In analogy with this latter case, when a finite concentration $x$ of magnetic impurities is embedded within 
a bath of electrons, the impurity electrons can communicate via the bath, giving origin to {\it Shiba bands}, 
which also appear within the superconducting gap. 

This is indeed what we find in our calculation, as shown in Fig.~\ref{STM1}(a), where we display 
the electronic density of states $N_{\uparrow/\downarrow}(\omega)=-\mathrm{Im}[G_{m \uparrow/\downarrow}(\omega)]/\pi$ at 
magnetic sites in the case with $x=0.01$. An odd-$\omega$ component also appears in non-magnetic sites by an inverse 
proximity effect, as the odd-$\omega$ superconductivity spreads throughout the system. 
In those sites, however, the amplitude is much smaller (see Appendix~\ref{nmsit} for an example). Therefore 
in the following we focus on the magnetic sites, where it would be easier to detect the odd-$\omega$ contribution.
Two YSR impurity bands are clearly visible within the superconducting gap, as displayed in the inset of Fig.~\ref{STM1}(a).
With this choice of parameters, the lower ($\omega < 0$) and upper ($\omega>0$) YSR bands are well separated and have spin down 
(dashed line) and spin up (solid line) character. We will discuss later the more complicated case 
where the lower and upper YSR bands overlap close to $\omega=0$. 
A possible experimental realization of this has been discussed in conjunction 
with the presence of Majorana fermions, which appear as a resonance in the density of states at
$\omega=0$~\cite{PergeSci,jeon2017distinguishing}. 


In contrast to dilute magnetic semiconductors, our system is in a superconducting state. We expect therefore superconductivity 
to be induced at the impurity site by proximity effect. Because of the time reversal symmetry breaking due to
the magnetic field at the impurity, an odd-$\omega$ pairing component is expected to appear\cite{Linder},
similarly to the single impurity case \cite{kuzmanovski2020odd,perrin2019unveiling}.
As the impurity electrons form the YSR bands, these should then display an odd-$\omega$ superconducting component in the superconducting function. 
In fact, such a component can be seen in Figs.~\ref{STM1}(b)-(c),
where we display the real and imaginary parts 
of the full superconducting function, $\mathrm{Re}[F_m(\omega)]$ and $\mathrm{Im}[F_m(\omega)]$, respectively. 
In a non-magnetic spin-singlet BCS-like superconductor $\mathrm{Re}[F_m(\omega)]$ is symmetric in $\omega$, 
whereas here it is not, in close analogy to what it is found in the single-impurity case \cite{kuzmanovski2020odd,perrin2019unveiling}. 
It is convenient to decompose the total  $F_m(\omega)$ in a standard even-$\omega$ $F_m^e$ and an odd-$\omega$ $F_m^o$ components, defined by $F_m^{e/o}(\omega)=\frac{F_m(\omega)\pm F_m(-\omega)^*}{2}$, which are also displayed in Figs.~\ref{STM1}(b)-(c). The function $F_m^{o}(\omega)$ describes triplet, $s$-wave, odd-$\omega$ pairing.

While the density of states (Fig.~\ref{STM1}(a)) can be directly obtained 
from STS measurements, it is in general difficult
to extract the superconducting function and, in particular, the odd-$\omega$ part (Figs.~\ref{STM1}(b)-(c)). 
Recently it was shown that this can be achieved in the case of a single isolated impurity, 
where the odd-$\omega$ Cooper pairs are localized at the impurity site \cite{perrin2019unveiling}.
We will now show, that this idea can be extended to 
the present case of delocalized YSR bands with odd-$\omega$ superconductivity. 
We will provide an explicit protocol to extract the superconducting function from STS experimental data. 
To this purpose, we will use the magnetic disorder concentration $x$ as a tuning scaling parameter and derive explicit expressions relating the STS density of states to the odd-$\omega$ superconducting funtion.

\subsection{YSR-band scaling with impurity concentration\label{3a}}
\begin{figure}[t] 
\includegraphics[scale=0.6]{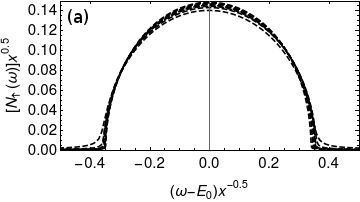}
 \includegraphics[scale=0.6]{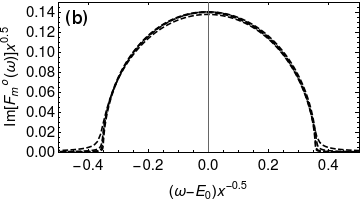}
\caption{Plots of Shiba bands centered at $E_0>0$ using the same parameters as in Fig.~\ref{STM1}, but with $\eta=10^{-4}$ and making a scaling with $x$. The horizontal axes are multiplied by $x^{-1/2}$ and the vertical ones by $x^{1/2}$. (a) shows plots of $N_\uparrow$ using $x=0.0001$, $0.001$, $0.005$, $0.01$, $0.015$, $0.02$, $0.025$, $0.03$.  
(b)~shows plots of $\mathrm{Im}[F_m^o(\omega)]$ using $x=0.0001$, $0.001$, $0.005$, $0.01$. The YSR bands overlap for $x>0.01$ and $\mathrm{Im}[F_m^o(\omega)]$ becomes the superposition of two semicircles. For this reason curves with higher values of $x$ were omitted in (b). It was important to consider a small broadening $\eta$ in order to cause the lines to overlap completely.
}\label{scaling}
\end{figure}
We first notice that the shape of the YSR bands (inset of Fig.~\ref{STM1}(a)) reflects the semicircle shape of the Bethe lattice density of states $D(\varepsilon)$. 
The presence of a quasiparticle band, with a renormalized mass and a density of states that reflects
the non-interacting one of the lattice, is a well-known feature of the DMFT solution for strongly interacting metallic states
~\cite{Rozenberg1994}. 
Here, the strongly correlated states are those of the magnetic impurity network, whose states are subject to the local 
interaction $J$ that reduces their effective hopping. Thus, we may expect that this quasiparticle heavy-band may carry the
information of the impurity concentration $x$.  
Thus, we may attempt a scaling of the YSR bands as a 
function of the disorder-site density $x$.
In Fig.~\ref{scaling}(a) we plot the upper YSR band for various values of $x$ scaled as
\begin{eqnarray}
 N_{\uparrow/\downarrow}(\omega)&=& \frac{a_{\uparrow/\downarrow}}{bx^{1/2}}D\left(\frac{\omega\mp E_0}{bx^{1/2}}\right),\label{fitN}
\end{eqnarray}
where $E_0,~ b,~a_\uparrow$ and $a_\downarrow$ depend on the model parameters $t$, $\mu$, $\delta_\mu$, $\Delta$ and $J$, but do not change significantly with $x$. Note that $\omega=\pm E_0$ mark the midpoints of the YSR bands.
Also notice that $a_{\uparrow/\downarrow}$ have a priori different values for each of the two Shiba bands.
Magnetic disorder acts to rescale the width and the height of the YSR bands according to a $\sqrt{x}$ dependence.
The collapse of all the curves on the same line proves the validity of such a scaling.  

A similar scaling applies to the superconducting function as well. In Fig.~\ref{scaling}(b) 
we show that the odd-$\omega$ component $\mathrm{Im}[F_m^o(\omega)]$ has the same shape of $D(\omega)$ 
and the same scaling with $x$,
\begin{eqnarray}
  \mathrm{Im}[F_m^o(\omega)]&=&\frac{a_{F}}{bx^{1/2}} \sum_{s=\pm 1}D\left(\frac{\omega-s E_0}{bx^{1/2}}\right).
  \label{fitF}
\end{eqnarray}
These results lead us to establish a useful relation between the odd-$\omega$ superconducting function
and the density of states
\begin{eqnarray}
 \mathrm{Im}[F_m^o(\omega)] &=& a_F\left[\frac{N_{\uparrow}(\omega)}{a_\uparrow}+ \frac{N_{\downarrow}(\omega)}{a_\downarrow}\right],\label{fitFN}
\end{eqnarray}
This relation is similar to the one obtained for the single impurity case \cite{perrin2019unveiling}, as we show explicitly in Sec.~\ref{deriv}. However its range of validity is now extended to
superconductors dirtied with a low concentration of magnetic impurities. 
Equations (\ref{fitN})-(\ref{fitFN}) remain valid even when the Shiba bands overlap (see Sec.~\ref{deriv}).
\begin{figure}[tb!] 
 \includegraphics[scale=0.65]{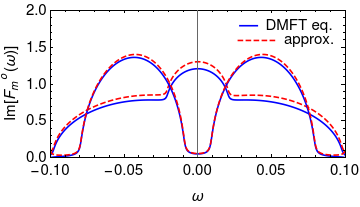}
\caption{$\mathrm{Im}[F_m^o(\omega)]$ obtained using the same parameters as in Fig.~\ref{STM1} and $\eta=10^{-3}$, but with different values of $x$, $x=0.01$, where the two Shiba bands do not overlap, and $x=0.03$, where the two bands do overlap, giving rise to a resonance around $\omega=0$. Solid blue lines were obtained using the DMFT equations (\ref{dmfteq22}), while dashed red lines correspond to Eq.~(\ref{fitFN4}).}\label{ImFo}
\end{figure}

STS measurements of the spin-polarized density of states could allow us, via Eq.~(\ref{fitFN}), to 
extract the odd-$\omega$ component of the superconducting function, provided the coefficient $a_F$ is determined.
This cannot be obtained directly from a spectroscopic measurement.
We show however in the following section that a relation between 
$a_{\uparrow/\downarrow}$ and $a_F$ can be explicitly derived by going to the single impurity limit,
as these coefficients depend only weakly on the impurity concentration $x$. If the normal-state 
density of states at the Fermi level can be assumed to be constant within the superconducting gap
(i.e.~particle-hole symmetry can be assumed
at low energy), which is the case of most standard superconductors, we will show that
\begin{eqnarray}
\label{cFud}
 a_F=\mathrm{sgn}(J\Delta)\frac{\pi}{2}\sqrt{a_\uparrow a_\downarrow}.\label{cfud}
\end{eqnarray}
With this relation we find the following expression that relates the
odd-frequency superconducting function directly to the spin-polarized density of states $N_{\uparrow/\downarrow}(\omega)$,
which are \textit{a priori} directly measurable functions: 
\begin{eqnarray}
  \mathrm{Im}[F_m^o(\omega)] &=& \mathrm{sgn}(J\Delta)\frac{\pi}{2}\sqrt{N_{\uparrow}(E_0) N_{\downarrow}(-E_0)} \nonumber\\
  & \times & \bigg[\frac{N_{\uparrow}(\omega)}{N_{\uparrow}(E_0)}
  + \frac{N_{\downarrow}(\omega)}{N_{\downarrow}(-E_0)}\bigg].\label{fitFN4}
\end{eqnarray}

Notice that in most cases, when the lower and upper YSR bands do not overlap significantly at $\omega=0$, the
YSR  bands are fully spin polarized (as in Fig.~\ref{STM1}(a)). We can in this case identify
$N_{\uparrow}(\omega)= N(\omega> 0)$ and $N_{\downarrow}(\omega)= N(\omega< 0)$, where
$N(\omega)= N_{\uparrow}(\omega)+ N_{\downarrow}(\omega)$ is the total density of states which is more easily accessible with
standard STS. We can then simply write
\begin{eqnarray}
  \mathrm{Im}[F_m^o(\omega)] &=& \mathrm{sgn}(J\Delta)\frac{\pi}{2}\sqrt{N_{}(E_0) N_{}(-E_0)} \nonumber \\
  & \times & \bigg[\frac{N_{}(\omega)}{N_{}\big(\mathrm{sgn}(\omega) E_0\big)} \bigg].\label{fitFN5}
\end{eqnarray}
Furthermore, even when the lower and upper YSR bands do overlap at $\omega=0$, it would still be
possible to attempt a fit of the experimental density of states which could separate the 
$N_{\uparrow/\downarrow}(\omega)$ components.
Then, Eq.~(\ref{fitFN4}) can still be used as a good approximation if the YSR bands overlap over a short interval inside the superconducting gap.
Finally, when even this last case is not applicable, it would still be possible to measure $N_{\uparrow/\downarrow}(\omega)$ components by employing spin-dependent STS \cite{Wiesendanger09}.
From the experimental point of view, it could in principle be difficult to reproduce the concentration scaling
  analysis that we carried
  out theoretically to derive Eq.~(\ref{fitFN}),
  as it is well known that an increasing concentration of magnetic impurities rapidly
  suppresses superconductivity. The key relations~(\ref{fitFN})-(\ref{fitFN5}) remain valid however at any small given $x$
  and can be used in experiments to extract the odd-$\omega$ component $\mathrm{Im}[F_m^o(\omega)]$. 

We display in Fig.~\ref{ImFo}  the $\mathrm{Im}[F_m^o(\omega)]$ theoretically extracted using Eq.~(\ref{fitFN4})
for different values 
of the model parameters, including a case where the YSR bands overlap close to $\omega=0$. The curves (dashed lines) 
are compared with the $\mathrm{Im}[F_m^o(\omega)]$ obtained directly from the solution of the DMFT equations 
(\ref{dmfteq22}) (solid lines). 
The good agreement is a proof of principle of our analysis and validates our results. 
We finally remark that once $\mathrm{Im}[F_m^o(\omega)]$ is extracted using the method presented here, $\mathrm{Re}[F_m^o(\omega)]$ can be obtained via the Kramers-Kronig relations.

\subsection{Derivation of relations\label{deriv}}
We will now prove that the relation established from the disorder concentration scaling, Eq.~(\ref{fitFN}),
can be justified analytically 
and we derive Eq.~(\ref{cFud}) explicitly by going to the well controlled single impurity limit.   

In the small concentration regime $x \ll 1$, magnetic impurities  affect only mildly the bulk superconducting Green's function $\hat{G}_{s}$.
We can then assume 
$
\hat{G}_{nm}\simeq \hat{G}_{s}.
$
As $\hat{G}_{m}^{-1}=\hat{G}_{nm}^{-1}+\hat{R}$, where $\hat{R}=\delta_\mu\tau^z+\Delta\tau^x-J\openone$ (see Eqs.~\ref{dmfteq22}), we can write $ \hat{G}_{m}^{-1}=\hat{G}_{s}^{-1}+\hat{R}.$
From this relation we derive the equations (see Appendix~\ref{appa}),
\begin{eqnarray}
   N_{\uparrow}(\omega)\pm N_{\downarrow}(\omega)&=&-\frac{2}{\pi}\mathrm{Im}\bigg[\frac{G_s(\omega)-\delta_\mu \texttt{det}\hat{G}_s(\omega)
   }{F_s(\omega)-\Delta\texttt{det}\hat{G}_s(\omega)}F_m^{e/o}(\omega)\nonumber\\
    & &- \frac{J\texttt{det}\hat{G}_s(\omega)}{F_s(\omega)-\Delta\texttt{det}\hat{G}_s(\omega)}F_m^{o/e}(\omega)\bigg],\label{UpmD}
\end{eqnarray}
where $G_s(\omega),F_s(\omega)$ are elements of the matrix $\hat{G}_s(\omega)$.
These expressions illustrate well the DMFT results, as we show in Appendix~\ref{appa}.

Here we follow Ref.~\onlinecite{perrin2019unveiling}, generalizing it to the case of many magnetic impurities, to obtain some useful expressions for low energies. 
For $|\omega|<|\Delta|$, inside the superconducting gap, we can approximate the denominator of Eq.~(\ref{UpmD}) by $F_s(\omega)-\Delta\texttt{det}\hat{G}_s(\omega)\simeq F_s(\omega)$ if $|\Delta/t|\ll1$, which is the case of conventional superconductors. 
From the definition $F_m^{e/o}(\omega)=\frac{F_m(\omega)\pm F_m(-\omega)^*}{2}$, we find 
\begin{equation}
\mathrm{Im}\{F_m^e(\omega)-\mathrm{sgn}(\omega E_0)[F_m^o(\omega)-F_m(-\mathrm{sgn}(E_0)|\omega|)]\}=0.
\end{equation}
Multiplying this equality by $(G_s(\omega)-G_s^*(-\omega))/[\pi F_s(\omega)]$ and adding the result to Eq.~(\ref{UpmD}), we find
\begin{eqnarray}
   N_{\uparrow}(\omega)+ N_{\downarrow}(\omega)
     &=& C_o(\omega)\mathrm{Im}[F_m^{e}(\omega)]+C_e(\omega)\mathrm{Im}[F_m^o(\omega)] \nonumber \\
     & & +C_r(\omega)\mathrm{Im}[F_m(-\mathrm{sgn}(E_0)|\omega|)],\label{UpDVF}
\end{eqnarray}
where
\begin{eqnarray}
      C_o(\omega)&=& -\frac{1}{\pi}
      \frac{G_s(\omega)+ G_s^*(-
\omega)}{F_s(\omega)}+\frac{2\delta_\mu}{\pi} \frac{\texttt{det}\hat{G}_s(\omega)}{F_s(\omega)}, \nonumber \\
      C_e(\omega)&=&
      \frac{2J}{\pi} \frac{\texttt{det}\hat{G}_s(\omega)}{F_s(\omega)}-
       \frac{\mathrm{sgn}(\omega E_0)}{\pi} \frac{G_s(\omega)- G_s^*(-\omega)}{F_s(\omega)}, \nonumber \\
        C_r(\omega)&=&\frac{\mathrm{sgn}(\omega E_0)}{\pi}\frac{G_s(\omega)-G_s^*(-\omega)}{F_s(\omega)}.\label{Cee}
 \end{eqnarray}
Here we assumed the same density of states for both spin species in the clean superconductor.
%

These functions can be simplified for $\omega$ inside the gap, 
where the imaginary components of $G_s(\omega)$ and $F_s(\omega)$ vanish. 
For a given lattice density of states $D(\epsilon)$, these coefficients can be expressed in terms of 
model parameters (see Appendix~\ref{SGeneral}), by making a power series expansion of $D(\epsilon)$ around $\epsilon=\mu$ 
and keeping the first order terms. We find
\begin{eqnarray}
      C_o(\omega)&\simeq& \left[-\frac{4}{\pi^2} 
      \frac{ D'(\mu) W }{ D(\mu) } +2\delta_\mu D(\mu) \right] \frac{\sqrt{\Delta^2-\omega^2}}{\Delta}, \nonumber \\
      C_e(\omega)&\simeq&
      2J \,  D(\mu)  \, \frac{\sqrt{\Delta^2-\omega^2}}{ \Delta }-
      \mathrm{sgn}(E_0)\frac{2}{\pi} \frac{|\omega|}{\Delta},
      \nonumber \\
      C_r(\omega)&\simeq &\mathrm{sgn}(E_0)\frac{2}{\pi} \frac{|\omega|}{\Delta}.\label{Cee1}
 \end{eqnarray}
Here $W$ is a cutoff proportional to the bandwidth and 
 $D'(\epsilon)$ is the derivative of the density of states.  
%
The function $\mathrm{Im}[F_m(\omega)]$ vanishes outside its resonance around $\omega=E_0$ (see e.g.~Fig.~\ref{STM1}(c)) and hence the function $\mathrm{Im}[F_m(-\mathrm{sgn}(E_0)|\omega|)]$ that appears in Eq.~(\ref{UpDVF}) vanishes for every $\omega$ where the YSR bands do not overlap. If $x$ is small enough, these bands are expected to overlap only for $|\omega|\ll |\Delta|$, where $C_r(\omega)$ is negligible (see Eq.~(\ref{Cee1})). Therefore the last term of Eq.~(\ref{UpDVF}) will be neglected.

The function $\mathrm{Im}[F_m^{e}(\omega)]$ is antisymmetric while $\mathrm{Im}[F_m^{o}(\omega)],~ C_e(\omega)$ and $C_o(\omega)$ are symmetric 
with respect to $\omega$.
Extracting the symmetric ($s$) and antisymmetric~($a$) components of Eq.~(\ref{UpDVF}) leads to
  \begin{eqnarray}
 [N_{\uparrow}(\omega)+ N_{\downarrow}(\omega)]^{s/a}  & \simeq & C_{e/o}(\omega) \mathrm{Im}[F_m^{o/e}(\omega)]. \label{tempss}
 \end{eqnarray} 
%
We now notice that the functions $N(\omega)= N_{\uparrow}(\omega)+ N_{\downarrow}(\omega)$ and $\mathrm{Im}[F_m^{o/e}(\omega)]$  vanish quickly as 
one moves away from $\omega= E_0$, while $C_{e/o}(\omega)$ vary slowly near $\omega= E_0$.
Therefore, we can safely replace $C_{e/o}(\omega)$ with $C_{e/o}(E_0)$:
\begin{eqnarray}
 [N_{\uparrow}(\omega)+ N_{\downarrow}(\omega)]^{s/a}  & \simeq & C_{e/o}(E_0) \mathrm{Im}[F_m^{o/e}(\omega)].\label{temps0}
\end{eqnarray} 
Equation~(\ref{temps0}) is then a good approximation even when the YSR bands overlap with one another, if the overlap takes place over a short interval inside the gap.

We now derive the relation between the odd-$\omega$ superconducting function and the density of states, Eq.~(\ref{fitFN}), 
that we inferred from the disorder concentration scaling analysis of the YSR bands. We first notice 
that integrating Eq.~(\ref{fitN}) one finds $\int_{-\infty}^{+\infty}N_{\uparrow/\downarrow}=a_{\uparrow/\downarrow}$ and
then $N_\uparrow(-\omega)/a_\uparrow=N_\downarrow(\omega)/a_\downarrow$. Plugging these relations into Eq.~(\ref{temps0})
one recovers Eq.~(\ref{fitFN}) 
provided that
\begin{eqnarray}
a_F= \, \frac{ a_\uparrow + a_\downarrow }{ 2 C_e(E_0)}. 
\label{cF1}
\end{eqnarray}
Note that $C_e(E_0)$ is given by Eq.~(\ref{Cee1}) and in the particular case where $|E_0| \ll |\Delta|$ one has $C_e(E_0)\simeq \mathrm{sgn}(\Delta)2J D(\mu)$.
Via the relation (\ref{fitFN}) we can finally extract the odd-$\omega$ superconducting
function, if we can determine the magnetic coupling $J$. As the latter 
may be difficult to extract from experiments, we can further simplify Eq.~(\ref{cF1}) by going to the single impurity 
limit. 

As we have mentioned in the disorder concentration 
scaling analysis, $a_{\uparrow/\downarrow}$ and $a_F$ do not depend on $x$. 
They thus remain the same 
in the single impurity limit, $x\rightarrow 0$, and Eq.~(\ref{fitN}) becomes a delta-like Lorentzian,
\begin{eqnarray}
N_{\uparrow/\downarrow}=\frac{\eta a_{\uparrow/\downarrow}/\pi}{(\omega\mp E_0)^2+\eta^2},\label{impLor}
\end{eqnarray}
where we introduced a finite inverse lifetime parameter $\eta$ for the Shiba state, replacing $\omega \to \omega+ i \eta$. 
Similarly, Eq.~(\ref{fitFN}) can be used together with Eq.~(\ref{impLor}) to find a Lorentzian expression for $\mathrm{Im}[F_m^o(\omega)]$:
\begin{eqnarray}
\mathrm{Im}[F_m^o(\omega)]&=&\sum_{s=\pm}\frac{\eta  a_F/\pi}{(\omega-s E_0)^2+\eta^2}.\label{impLorF}
\end{eqnarray}
%

On the other hand, we can take the DMFT equations~(\ref{dmfteq22}) in the $x\to 0$ limit
(see Appendix~\ref{SILC} for details). Once again we 
replace $\hat{G}_{nm}\simeq \hat{G}_{s}$, and obtain the Lorentzian 
form of $N_{\uparrow/\downarrow}$ and $\mathrm{Im}[F_m^o(\omega)]$ as a function
of the model parameters $D(\omega)$, $J$, $ \Delta $, $ \mu $, $\delta_\mu$ (see Eq.~(\ref{impclass})), 
provided that  
\begin{eqnarray}
a_{\uparrow/\downarrow}&=&\frac{2\pi D(\mu)|\alpha\Delta|[\gamma+2\alpha^2\pm \alpha( -h_A +2\beta)]}{(\gamma^2+4\alpha^2)^{3/2}}, \label{coefud}\\
a_F&=&\frac{\pi^2 D(\mu)\alpha\Delta}{\gamma^2+4\alpha^2},\label{coeff}
\end{eqnarray}
where we define $\alpha=\pi D(\mu)J$, $\beta=\pi D(\mu)\delta_\mu, \gamma=1-h_A\beta +\beta^2-\alpha^2$ and $h_A= 4 D^{\prime}(\mu) W/[\pi D(\mu)] \ll 1$, i.e.~we assume the density of states is almost constant at low energies.
Then we finally obtain
\begin{eqnarray}
a_F&=&\mathrm{sgn}(\alpha\Delta)\frac{\pi}{2}\sqrt{a_\uparrow a_\downarrow},\label{cff}
\end{eqnarray} 
which proves the relation (\ref{cFud}). 

\section{Conclusion\label{conclu}}

We have shown that odd-$\omega$ superconductivity occurs in a dilute magnetic superconductor model, 
where a finite concentration of magnetic impurities is randomly embedded in a conventional superconductor.
We have solved this model by treating the coupling between magnetic impurities and
superconducting electrons in dynamical mean field theory. 
This method is also appropriate to deal with the disorder 
and, within our approximations, is equivalent to the coherent potential approximation. This technique allows us to have
direct access to the local spectral functions that can be experimentally extracted by scanning tunneling experiments.
Our results show the formation of Yu-Shiba-Rusinov bands inside the bulk superconducting gap displaying a relevant odd-frequency
component. We have analyzed the YSR bands by means of a scaling analysis and derived 
an expression (Eq.~(\ref{fitFN4})) that allows us to explicitly extract the odd-frequency 
superconducting function from spectroscopic quantities that are directly accessible 
in scanning tunneling measurements.
To get this explicit formula we have exploited the findings of our disorder-concentration 
scaling by going to the single impurity limit, where exact expressions could be derived within physically 
reasonable approximations. We benchmarked our approximate expressions and 
found a very good correspondence with the dynamical mean field theory results, 
providing further proof of the validity of our analysis.

Our results should motivate future experimental investigations in dilute magnetic superconductors to search
for odd-frequency superconductivity. In particular, the formation of YSR bands possessing odd-frequency
character, which delocalize into the bulk superconductor, should raise further interesting
questions about the impact of the
odd-frequency pairing not only on the spectral properties but also on other thermodynamic and transport
properties, with an eye for future spintronic device applications.

\begin{acknowledgments}
This work was supported by FAPEMIG (F.L.N.S. and
M.C.O.A.); CNPq through Grants No. 142153/2016-8
(F.L.N.S.), No. 304983/2017-9 (M.C.O.A.),
No. 307041/2017-4 (E.M.), and No. CNPq INCT-IQ
465469/2014-0 (F.L.N.S. and M.C.O.A.); and CAPES, in
particular through programs CAPES-COFECUB-0899/2018
(F.L.N.S., M.C., P.S., M.C.O.A., E.M., and M.J.R.) and
CAPES-PrInt-UFMG (M.C.O.A.).
\end{acknowledgments}


\appendix

\section{Results for non-magnetic sites\label{nmsit}}

\begin{figure}[t] 
 \includegraphics[scale=0.62]{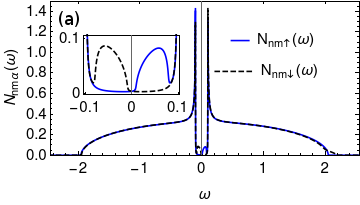}
 \includegraphics[scale=0.65]{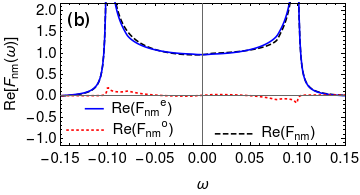}
 \includegraphics[scale=0.65]{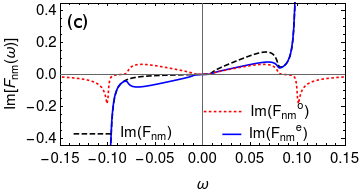}
\caption{Numerical results for non-magnetic sites obtained using the DMFT equations (\ref{dmfteq22}) with $t=1,~ \mu=-0.05,~\delta_\mu=-0.5,~ \Delta=-0.1,~J=-0.65,~ x=0.01$ and broadening $\eta=10^{-3}$. In panel (a) we show the full spectrum for $N_{nm\uparrow}$ and $N_{nm\downarrow}$;  in the inset we see these functions at low energies, $|\omega|<|\Delta|$. In (b) and (c) we present the real and imaginary components of the superconducting function $F_{nm}(\omega)$, respectively, as well as its even-$\omega$ and odd-$\omega$ components, at low energies.}\label{nmsites}
\end{figure}

On non-magnetic sites the YSR bands are much smaller than on magnetic ones. In Fig.~\ref{nmsites} we show the density of states on non-magnetic sites, $N_{nm\uparrow/\downarrow}(\omega)=-\mathrm{Im}[G_{nm \uparrow/\downarrow}(\omega)]/\pi$, and the pairing function $F_{nm}(\omega)$.

\section{Approximation for the Density of States on magnetic sites\label{appa}}

\begin{figure}[t] 
\includegraphics[scale=0.6]{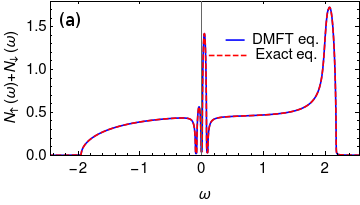}
\includegraphics[scale=0.6]{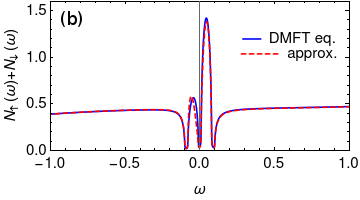}
\caption{Plots of the density of states, $N_\uparrow(\omega)+N_\downarrow(\omega)$, using $t=1,~ \mu=-0.05,~\delta_\mu=-0.5,~ \Delta=-0.1,~J=-0.65,~ x=0.01$ and $\eta=10^{-3}$. Solid blue lines give the solution of the DMFT equations (\ref{dmfteq22}), as presented in Fig.~\ref{STM1}. In (a) the dashed red line corresponds to Eq.~(\ref{exacteq}), showing that it is exact for the whole spectrum. In (b) the dashed red line corresponds to the approximate Eq.~(\ref{UpmD}), showing that the latter is a good approximation for the former for a broad range of energies.}\label{STM2}
\end{figure}

From the DMFT equations~(\ref{dmfteq22}) we can write 
\begin{eqnarray}
\hat{G}_{m}^{-1}=\hat{G}_{nm}^{-1}+\hat{R},\label{dmfteq33}
\end{eqnarray}  
where $\hat{R}=\delta_\mu\tau^z+\Delta\tau^x-J\openone$. Substituting the expression for the inverse of the $2\times 2$ matrices $\hat{G}_{m}$ and $\hat{G}_{nm}$ in Eq.~(\ref{dmfteq33}) we obtain the relations
\begin{eqnarray}
\frac{F_m(\omega)}{\texttt{det}\hat{G}_m(\omega)}&=&\frac{F_{nm}(\omega)}{\texttt{det}\hat{G}_{nm}(\omega)}-\Delta,\label{apA1}\\
\frac{G_{m\uparrow}(\omega)}{\texttt{det}\hat{G}_m(\omega)}&=&\frac{G_{nm\uparrow}(\omega)}{\texttt{det}\hat{G}_{nm}(\omega)}-J-\delta_\mu,\label{apA2}\\
-\frac{G_{m\downarrow}^*(-\omega)}{\texttt{det}\hat{G}_m(\omega)}&=&-\frac{G_{nm\downarrow}^*(-\omega)}{\texttt{det}\hat{G}_{nm}(\omega)}-J+\delta_\mu.\label{apA3}
\end{eqnarray}
From Eq.~(\ref{apA1}) we find an expression for $\texttt{det}\hat{G}_m(\omega)$ and using it in Eqs.~(\ref{apA2})-(\ref{apA3}) we find 
\begin{eqnarray}
N_\uparrow(\omega)\pm N_\downarrow(\omega)&=&-\frac{1}{\pi}\mathrm{Im} \bigg\{\frac{F_m(\omega)}{F_{nm}(\omega)-\Delta\texttt{det}\hat{G}_{nm}(\omega)}\nonumber\\
& &\times\left[ G_{nm\uparrow}(\omega)-(J+\delta_\mu)\texttt{det}\hat{G}_{nm}(\omega)\right]\nonumber\\
& & \pm \frac{F_m^*(-\omega)}{F_{nm}^*(-\omega)-\Delta\texttt{det}\hat{G}_{nm}^*(-\omega)}\nonumber\\
& &\times\Big[ G_{nm\downarrow}(\omega)+(J-\delta_\mu)\nonumber\\ & &\times \texttt{det}\hat{G}_{nm}^*(-\omega)\Big]\bigg\}. \label{exacteq}
\end{eqnarray}
Note that Eq.~(\ref{exacteq}) was obtained from the DMFT equations~(\ref{dmfteq22}) without any further approximation. 
Now assuming a low density of impurities, $x\ll 1$, we can write $\hat{G}_{nm}(\omega)\simeq \hat{G}_{s}(\omega)$, since $nm$ sites are only weakly affected by the impurities, and Eq.~(\ref{exacteq}) leads to Eq.~(\ref{UpmD}). 
Figure~\ref{STM2} compares $N_\uparrow+ N_\downarrow$ obtained numerically from the DMFT equations~(\ref{dmfteq22}) with the one obtained through Eq.~(\ref{exacteq}) in Fig.~\ref{STM2}(a), and with Eq.~(\ref{UpmD}) in Fig.~\ref{STM2}(b), where we use $G_s,~ F_s,~G_{nm},~F_{nm},~ F_m^e,~ F_m^o$ obtained numerically. We can see in the plots that Eq.~(\ref{UpmD}) is a good approximation to the DMFT solution for a broad range of energies. However this approximation might be inappropriate for high energies, $|\omega|\sim W$, where $\hat{G}_{nm}(\omega)$ may be more strongly affected by the impurities, and the approximation $\hat{G}_{nm}(\omega)\simeq \hat{G}_{s}(\omega)$ breaks down.

\section{Superconducting DMFT solution for general Density of States\label{SGeneral}}
%
The clean superconductor Green's function $G_s(i\omega_n)$ and anomalous Green's function $F_s(i\omega_n)$ can be computed using DMFT for general lattices \cite{Rozenberg},
\begin{eqnarray}
G_s(i\omega_n)&=&\int_{-\infty}^{+\infty} d\epsilon~ D(\epsilon)\frac{\zeta^*-\epsilon}{|\zeta-\epsilon|^2+\Sigma_\Delta^2},\\
F_s(i\omega_n)&=&-\Sigma_\Delta(i\omega_n)\int_{-\infty}^{+\infty} d\epsilon~ D(\epsilon)\frac{1}{|\zeta-\epsilon|^2+\Sigma_\Delta^2},~~~
\end{eqnarray}
where $D(\epsilon)$ is the density of states in the normal state, $\zeta=i\omega_n+\mu$ and we consider $\Sigma_\Delta\simeq\Delta\in \mathbb{R}$. 

Using analytical continuation and taking the imaginary part of $G_s$ and $F_s$, 
we obtain for $|\omega|>|\Delta|$,
\begin{eqnarray}
\mathrm{Im}[G_s(\omega)]
&=&\frac{\mathrm{sgn}(\omega)\pi}{2}
\Bigg\{\frac{-\omega}{\sqrt{\omega^2-\Delta^2}}\big[D(\mu +\sqrt{\omega^2-\Delta^2})\nonumber\\ & &+D(\mu - \sqrt{\omega^2-\Delta^2})\big]-\big[  D(\mu +\sqrt{\omega^2-\Delta^2})\nonumber\\
& &-D(\mu -\sqrt{\omega^2-\Delta^2})
\big]
\Bigg\},\nonumber\\ & &\\
 \mathrm{Im}[F_s(\omega)]&=&-\frac{\pi}{2}\mathrm{sgn}(\omega)\big[
D(\mu +\sqrt{\omega^2-\Delta^2})\nonumber\\ & &+D(\mu -\sqrt{\omega^2-\Delta^2}) \big]\frac{\Delta}{\sqrt{\omega^2-\Delta^2}}.
\end{eqnarray}
%
For $|\omega|<|\Delta|$, $\mathrm{Im}[G_s(\omega)]=\mathrm{Im}[F_s(\omega)]=0$.

Using the Kramers-Kronig relations and Taylor expanding $D(\mu \pm\sqrt{\omega^2-\Delta^2})$ up to first order around $\mu $, we find for $|\omega|<|\Delta|$
\begin{eqnarray}
 \mathrm{Re}[G_s(\omega)]&\simeq &
 -\pi D(\mu )\frac{\omega}{\sqrt{\Delta^2-\omega^2}}-2D'(\mu )W,\\
  \mathrm{Re}[F_s(\omega)]&\simeq &-\pi D(\mu )\frac{\Delta}{\sqrt{\Delta^2-\omega^2}},\label{FFF}
\end{eqnarray}
where $D'(\mu )=dD(\epsilon)/d\epsilon|_{\epsilon=\mu }$ and $W$ is a cutoff that corresponds to the edge of the energy band.

Considering the previous relations we can compute, inside the gap,
\begin{eqnarray}
\frac{G_s(\omega)+G_s^*(-\omega)}{F_s(\omega)}&\simeq &\frac{4D'(\mu )W}{\pi D(\mu )\Delta}\sqrt{\Delta^2-\omega^2},\label{GsFs}\\
\frac{G_s(\omega)-G_s^*(-\omega)}{F_s(\omega)}
&\simeq &
\frac{2\omega}{\Delta},\label{GsmFs}\\
\frac{\texttt{det}(\hat{G}_s(\omega))}{F_s(\omega)}&=&-\frac{G_s(\omega)G_s^*(-\omega)+F_s(\omega)^2}{F_s(\omega)}\nonumber\\
   &\simeq &\frac{\pi^2 D(\mu )^2+\left[2D'(\mu )W
 \right]^2}{\pi D(\mu )\Delta}
 \sqrt{\Delta^2-\omega^2}\nonumber\\
 &\simeq & \frac{\pi D(\mu)}{\Delta}\sqrt{\Delta^2-\omega^2},\label{detGF}
\end{eqnarray}
where we considered that the bands are almost flat, $D'(\mu)W\ll D(\mu)$.
Substituting the results given above in Eq.~(\ref{Cee}) we arrive at Eq. (\ref{Cee1}).


\section{Single impurity limiting case and Lorentzian functions\label{SILC}}

The semicircle described in Fig.~(\ref{scaling}) was obtained
making the substitution $\omega\rightarrow \omega +i \eta$,
  with $\eta=10^{-4}\simeq 0^+$.
In this case, we observe that the Green's function around a Shiba band has the form 
\begin{eqnarray}
G_{m\uparrow/\downarrow}(\omega)=\pi f_{\uparrow/\downarrow}\omega+d_{\uparrow/\downarrow}-i\pi \frac{a_{\uparrow/\downarrow}}{bx^{1/2}}D\left(\frac{\omega\mp E_0}{bx^{1/2}}\right)
,\nonumber\\
\end{eqnarray}
with additional parameters $f_{\uparrow/\downarrow}$ and $d_{\uparrow/\downarrow}$.
We can find an expression for $N_{\uparrow/\downarrow}(\omega)=-\mathrm{Im}[G_{m \uparrow/\downarrow}(\omega+i\eta)]/\pi$.
Requiring that $\lim_{\omega\rightarrow\pm \infty}N_{\uparrow/\downarrow}(\omega)=0$, we find
\begin{eqnarray}
 N_{\uparrow/\downarrow}(\omega)&=& \frac{a_{\uparrow/\downarrow}}{bx^{1/2}}\bigg\{\mathrm{Re}\bigg[D\left(\frac{\omega\mp E_0+i\eta}{bx^{1/2}}\right)\bigg]-\frac{g\eta}{bx^{1/2}}\bigg\}.\nonumber\\ 
 & &\label{fitN22}
\end{eqnarray}
For the Bethe lattice, $g=D(0)/W=1/(2\pi t^2)$, where $W=2t$ is a cutoff for the normal state band.

Equation (\ref{fitFN}) is valid for arbitrary values of $x$ and $\eta$ and we can use it together with Eq.~(\ref{fitN22}) to find 
\begin{eqnarray}
  \mathrm{Im}[F_m^o(\omega)]&=&\frac{a_F}{bx^{1/2}} \label{fitFN221}\\
  & \times & \sum_{s=\pm}\bigg\{\mathrm{Re}\bigg[ D\left(\frac{\omega-s E_0+i\eta}{bx^{1/2}}\right) \bigg]
 -\frac{g\eta}{bx^{1/2}} \bigg\}. \nonumber 
\end{eqnarray}

In the limit of a single impurity, $x\rightarrow 0$, Eqs.~(\ref{fitN22})-(\ref{fitFN221}) become the Lorentzian functions given by Eqs.~(\ref{impLor})-(\ref{impLorF}). The coefficients in these equations can be determined as follows.

From the DMFT equations for a single impurity, $x\to 0$, we have
\begin{eqnarray}
\hat{G}_{m}&=&[\hat{G}_{s}^{-1}+\hat{R}]^{-1}\nonumber\\
&=&\frac{1}{\texttt{det}[\hat{G}_s^{-1}(\omega)+\hat{R}]}
 \left(
   \begin{array}{cc}
g_{11} & g_{12}\\
g_{21} & g_{22}
   \end{array}\right),\nonumber\\ & & \label{gmm}
\end{eqnarray}
where
\begin{eqnarray}
g_{11}&=&\frac{G_s(\omega)}{\texttt{det}[\hat{G}_s(\omega)]}-J-\delta_\mu, \nonumber\\
g_{12}&=&g_{21}=\frac{F_s(\omega)}{\texttt{det}[\hat{G}_s(\omega)]}-\Delta,\nonumber\\
g_{22}&=&-\frac{G_s^*(-\omega)}{\texttt{det}[\hat{G}_s(\omega)]}-J+\delta_\mu.
\end{eqnarray}
We can safely consider $F_s(\omega)/\texttt{det}\hat{G}_s(\omega)-\Delta\simeq F_s(\omega)/\texttt{det}\hat{G}_s(\omega)$ for $|\omega|<|\Delta|$ when $|\Delta|/t\ll 1$.

After some algebra we find 
\begin{eqnarray}
\frac{1}{\texttt{det}[\hat{G}_s^{-1}(\omega)+\delta_\mu\tau^z-J\openone]}&=&\frac{\texttt{det}[\hat{G}_s(\omega)]}{F_s(\omega) \zeta(\omega)},
\end{eqnarray}
where
\begin{eqnarray}
\zeta(\omega)&=&(J^2-\delta_\mu^2)\frac{\texttt{det}(\hat{G}_s(\omega))}{F_s(\omega)}+\delta_\mu \frac{G_s(\omega)+G_s^*(-\omega)}{F_s(\omega)}\nonumber\\ 
& & -J \frac{G_s(\omega)-G_s^*(-\omega)}{F_s(\omega)}+ F_s^{-1}(\omega).\label{zetaa}
\end{eqnarray}
Notice that the functions that appear in Eq.~(\ref{zetaa}) are given explicitly in Eqs.~(\ref{FFF})-(\ref{detGF}).
Where the function $\zeta(\omega)$ vanishes a pole in the Green's function produces the YSR states within
  the gap. This occurs at $\omega=E_0$ given by
\begin{eqnarray}
E_0&=&-\texttt{sgn}(\alpha)\frac{|\Delta|\gamma}{\sqrt{\gamma^2+4\alpha^2}},\label{e00}
\end{eqnarray}
where $\gamma=1-h_A\beta +\beta^2-\alpha^2$.

We then include a broadening $\eta$ in the expressions (\ref{FFF})-(\ref{detGF}), substitute them in Eq.~(\ref{gmm}) and expand $\hat{G}_m(\omega)$ around $\omega=E_0$. Then we find the Lorentzian expressions in Eqs.~(\ref{impLor})-(\ref{impLorF}) with coefficients given in Eqs.~(\ref{coefud})-(\ref{coeff}).
We also rewrite $C_e(E_0)$ given by Eq.~(\ref{Cee1}) using Eq.~(\ref{e00}), and find
\begin{eqnarray}
C_e(E_0)&=&\frac{2\texttt{sgn}(\alpha \Delta)(\gamma+2\alpha^2)}{\pi \sqrt{\gamma^2+4\alpha^2}}\nonumber\\
&=&\frac{\texttt{sgn}(\alpha \Delta)(a_\uparrow+a_\downarrow)}{\pi \sqrt{a_\uparrow a_\downarrow}}.\label{ceee}
\end{eqnarray}

\bibliographystyle{apsrev4-1}
\bibliography{references}

\end{document}